# Long-Distance Signal Propagation in AC-LGAD

C. Bishop, A. Das, J. Ding, M. Gignac, F. Martinez-McKinney, S. M. Mazza, A. Molnar,
N. Nagel, M. Nizam, J. Ott, H. F.-W. Sadrozinski, B. Schumm, A. Seiden, T. Shin,
A. Summerell, M. Wilder, Y. Zhao

*SCIPP, Univ. of California Santa Cruz, CA 95064, USA*

*Abstract–* We investigate the signal propagation in AC-LGAD (aka RSD), which are LGAD with a common $N^+$ layer and segmented AC-coupled readout contacts, by measuring the response to IR laser TCT on a large selection of AC-LGAD with strip readout. The interest for this topic derives from the realization that while large charge sharing between neighboring strips is essential for good position resolution, large sharing beyond the next neighbor generates background signals which in general are detrimental to the sensor goal of low occupancy. Using AC-LGAD with strip readout produced by Hamamatsu Photonics (HPK), we evaluate the effects of a variety of sensor properties, including geometrical parameters (strip length, width), process parameters like the $N^+$ layer resistivity, the coupling capacitance, and the thickness of the bulk on the signal sharing and the position resolution.

PACS: 29.40.Gx, 29.40.Wk, 78.47jc

Keywords: fast silicon sensors; charge multiplication; AC-LGAD strips; charge sharing.

## 1. Introduction

Low-gain Avalanche Detectors (LGAD) have been recently introduced as fast semiconductor timing sensors [1,2]. In their large-scale experimental applications, the High Granularity Timing Detector (HGTD) in ATLAS [3] and the MIP Timing Detector (MTD) in CMS [4], their segmentation is limited to pads with about 1 mm pitch by consideration of power, fill-factor, and field uniformity. While the fill-factor and uniformity problems could be solved for small-scale applications with Trench-Isolated LGAD [5], an acceptable power density will limit the spatial resolution unless charge sharing between distant contacts is employed. This is the principle of the AC-LGAD technology (aka Resistive Silicon Detector RSD) [6-8] which is based on a complete integration of four of the sensor layers in common sheets of the P-type bulk, the $P^+$ gain layer, the $N^+$ layer and a dielectric sheet, separating the first three from the segmented metal readout contacts (Fig. 1). A signal originating in the bulk and amplified in the gain layer is then shared between several electronics channels, allowing reconstruction of signal location with a resolution which is a small fraction of the readout pitch. Yet due to the common $N^+$ layer, the observed signal in AC-LGADs is the sum of the directly induced signal from the movement of the collected charge on neighboring contacts and the pick-up ("leakage") of the signal conducted on the $N^+$ layer common to all contacts.

The relative strength between induced and conducted signal depends on a variety of sensor parameters which we compare in the following study using scanning laser Transient Current Technique (TCT) on strip AC-LGAD produced by Hamamatsu Photonics K.K. (HPK) [9]: the geometry of the metal readout contacts was varied, as were production details of two common layers ($N^+$ layer resistivity and dielectric specs) and the bulk thickness. The doping of the gain layer and the strip pitch were kept constant. This will allow to check the simple assumption that for a large local signal and a small amount of long-distance conducted signal a large $N^+$ resistivity and a large coupling capacitance are needed.

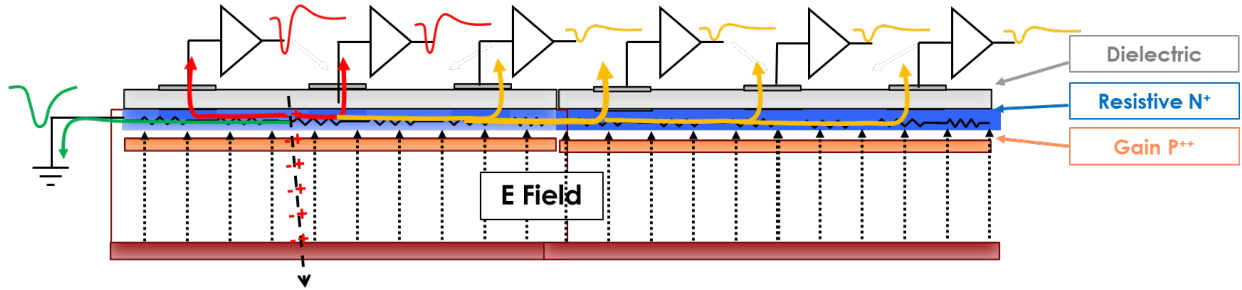

Fig. 1 Cross section of an AC-LGAD showing common sensor layers and the signals shared by neighboring metal contacts (red: direct induced signal, yellow: pick-up from the $N^+$ layer, i.e. "leakage") and the signal traveling in the $N^+$ layer and collected in the ground contact (green).

## 2. Experimental

### 2.1 Sensors

The sensors used were fabricated by HPK as part of the US-Japan Collaborative Agreement [10] and were partially funded by the eRD112 project to develop AC-LGAD detectors for the Electron-Ion Collider [11]. Figure 2 LEFT shows the layout of the approximately 5 mm wide AC-LGAD strip sensor with 5 mm long strips on 500 μm pitch and the wire bonds connecting 8 strips to the readout pads, with two strips left floating.

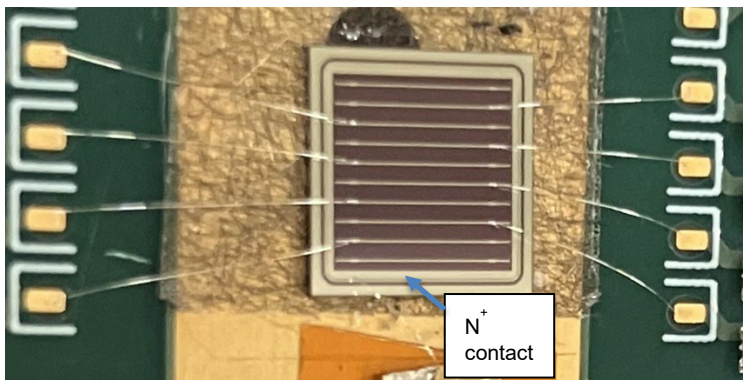 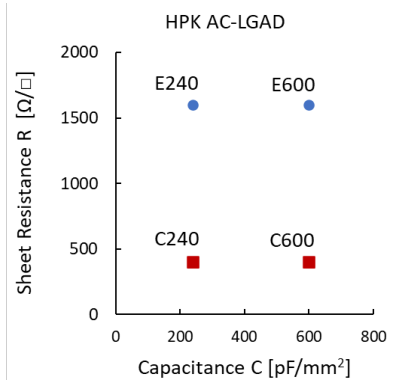

Fig. 2 LEFT: HPK AC-LGAD with 5 mm long strips on 500 μm pitch. Wire bonds between 8 strips and the readout and from the $N^+$ layer contact to ground are visible; RIGHT: Selected values for the sheet resistance of the $N^+$ layer and for the capacitance per area of the coupling capacitance [10].

Two values for the $N^+$ sheet resistance (called "E-Type" and "C-Type" in the following) and two values for the capacitance of the coupling capacitance were selected, resulting in four basic sensor combinations with the names shown in Fig. 2 RIGHT. The detailed sensor layout was then achieved by selecting two thicknesses of the bulk (50 and 20 µm), and for the metal strips, two widths (50 and 100 µm), and three length (5, 10, 20 mm) all on a 500 µm pitch as shown in Table 1. This allows to compare all four of the selections in Fig. 2 RIGHT. It should be pointed out that the actual value of the coupling capacitance varies between 60 pF and 1200 pF.

Table 1: Parameters of the tested HPK AC-LGAD

|  | Wafer | | | | Strip | | |
|---|---|---|---|---|---|---|---|
|  | Wafer # | $N^+$ Sheet Resistance [Ω/□] | Dielectric C (pF/mm$^2$) | Bulk Thickness T [µm] | Lenght L (mm) | Width W (µm) | Pitch P (µm) |
| HPK1 | W02 | E: 1600 | 240 | 50 | 5 | 50 | 500 |
| HPK3 | W05 | E: 1600 | 600 | 50 | 5 | 50 | 500 |
| HPK4 | W08 | C: 400 | 600 | 50 | 5 | 50 | 500 |
| HPK8 | W04 | C: 400 | 240 | 50 | 5 | 100 | 500 |
| HPK21 | W05 | E: 1600 | 600 | 50 | 10 | 100 | 500 |
| HPK22 | W08 | C: 400 | 600 | 50 | 10 | 100 | 500 |
| HPK27 | W05 | E: 1600 | 600 | 50 | 20 | 50 | 500 |
| HPK28 | W08 | C: 400 | 600 | 50 | 20 | 50 | 500 |
| HPK29 | W09 | E: 1600 | 600 | 20 | 20 | 50 | 500 |
| HPK35 | W09 | E: 1600 | 600 | 20 | 20 | 100 | 500 |

### 2.2 IR Laser TCT Measurements

The charge collection measurements using TCT follow the method described in [12]. In short, the sensors are mounted on fast analog amplifier boards with 16 channels and 1 GHz of bandwidth designed at Fermilab (FNAL) [13] and read out by a fast oscilloscope (2 GHz, 20 Gs). The sensors are excited with an infrared (IR) 1064 nm pulsed laser with a pulse temporal width of 400 ps, and a spot of 10-20 µm width mimicking the response of a MIP in the silicon [14]. The IR laser cannot penetrate the metal strips; therefore, the sensor behavior can be characterized only in between metal electrodes.

The read-out board is mounted on X-Y moving stages having close to 1 µm precision [15], so the response of the sensor as a function of laser illumination position can be evaluated. The scans are performed in 10 µm steps at a right angle to the strips at their midpoint, starting and ending at the $N^+$ contact. At each position, 100 waveforms are averaged to decrease the effect of laser power fluctuations, and a photodiode is used to correct for them. The scans are analyzed using the pulse shape in each position to derive the pulse maximum (Pmax) [12]. In addition, the rise time and fall time and the time of arrival for scans along and across the strips were recorded and will be the topic of a future publication.

### 3. Results and Discussion

#### 3.1 Pulse Height Pmax

An important parameter of the LGAD is the internal gain, which multiplies the generated pulse charge to the level of the collected charge. Yet most of the LGAD investigations of timing or location precision involve the pulse height instead of the charge. The maximum pulse height, Pmax, is shown in Figure 3 for all sensors for bias voltages below the on-set of a breakdown. The exponential gain dependence is similar for all sensors, but the curves separate according to being E-Type or C-Type. Sensors with longer strips require larger bias voltage to reach the same Pmax. Sensors with bulk thickness T = 20 μm and T = 50 μm show essentially the same Pmax range due to different weighting fields and rise times, yet at different bias voltages. With a noise value of N = 1 mV, the signal-to-noise ratio is S/N = 100 for Pmax = 100 mV.

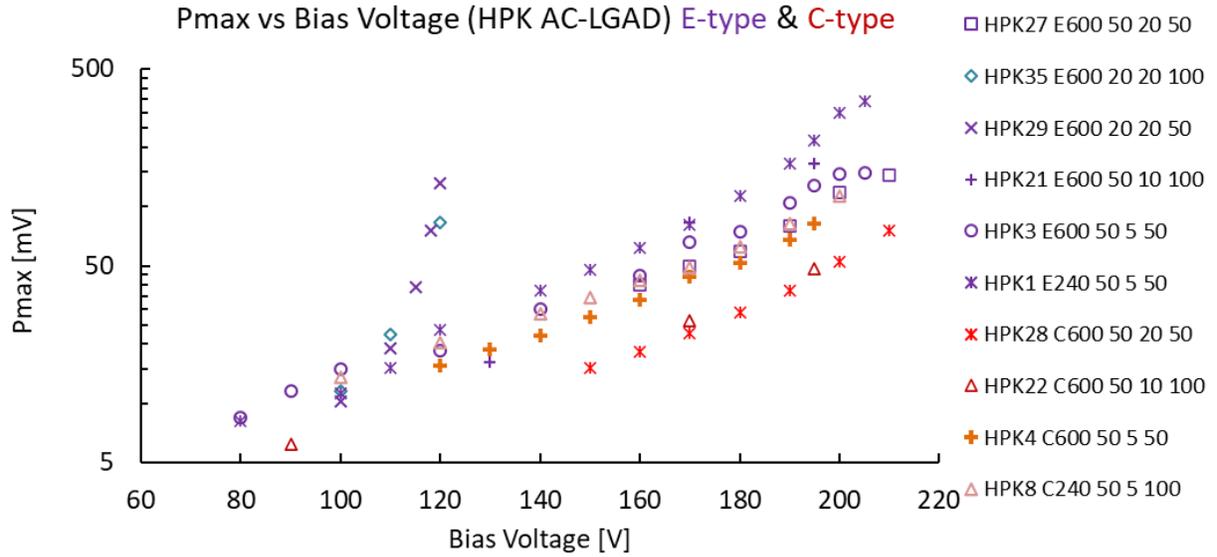

Fig. 3 Bias voltage dependence of the pulse height Pmax for the tested AC-LGADs (E-Type in purple and C-Type in red) (Legends: ID#, Type, Thickness T, Strip Length L, Strip Width W).

### 3.2 Pulse Sharing

The pulse sharing in the different AC-LGAD is tested with Pmax distributions from TCT laser scans across the strips. To allow comparisons of different sensors and bias conditions, Pmax gets normalized by dividing the measured values at each position by the maximum value found in a scan. The normalized Pmax distributions are shown in Fig. 4, with Fig. 4 TOP comparing the four sensor combinations indicated in Fig. 2 RIGHT for 5 mm long strips, Fig. 4 MIDDLE comparing sensors with the same oxide thickness but different strip length and $N^+$ resistivity, and Fig. 4 BOTTOM 20 mm long strips with different $N^+$ resistivity and/or bulk thickness (20 μm and 50 μm). (N.B.: some of the positional scans exhibit increases at the location of the two floating strips, being attributed to pick-up from the close-by $N^+$ layer contact). Different sensors show different signal sharing properties depending to the Type (Fig. 4 TOP):

- E600 and E240 (high R) have optimal close signal sharing contained within the strip center of the next neighbor, getting reduced to 2 – 3 % "leakage" at long distance, with E600 preferred,
- C600 (low R, high Cap) has large sharing close to and beyond the next neighboring strip, exhibiting long-distance constant "leakage" of the order of 10%,
- C240 (low R, low Cap) has large sharing to the next few neighbors, with the long-distance "leakage" reduced to < 2%.

This large sharing in the C-Type sensors appears to be the root cause for the reduced Pmax compared to E-Types shown in Fig. 3.

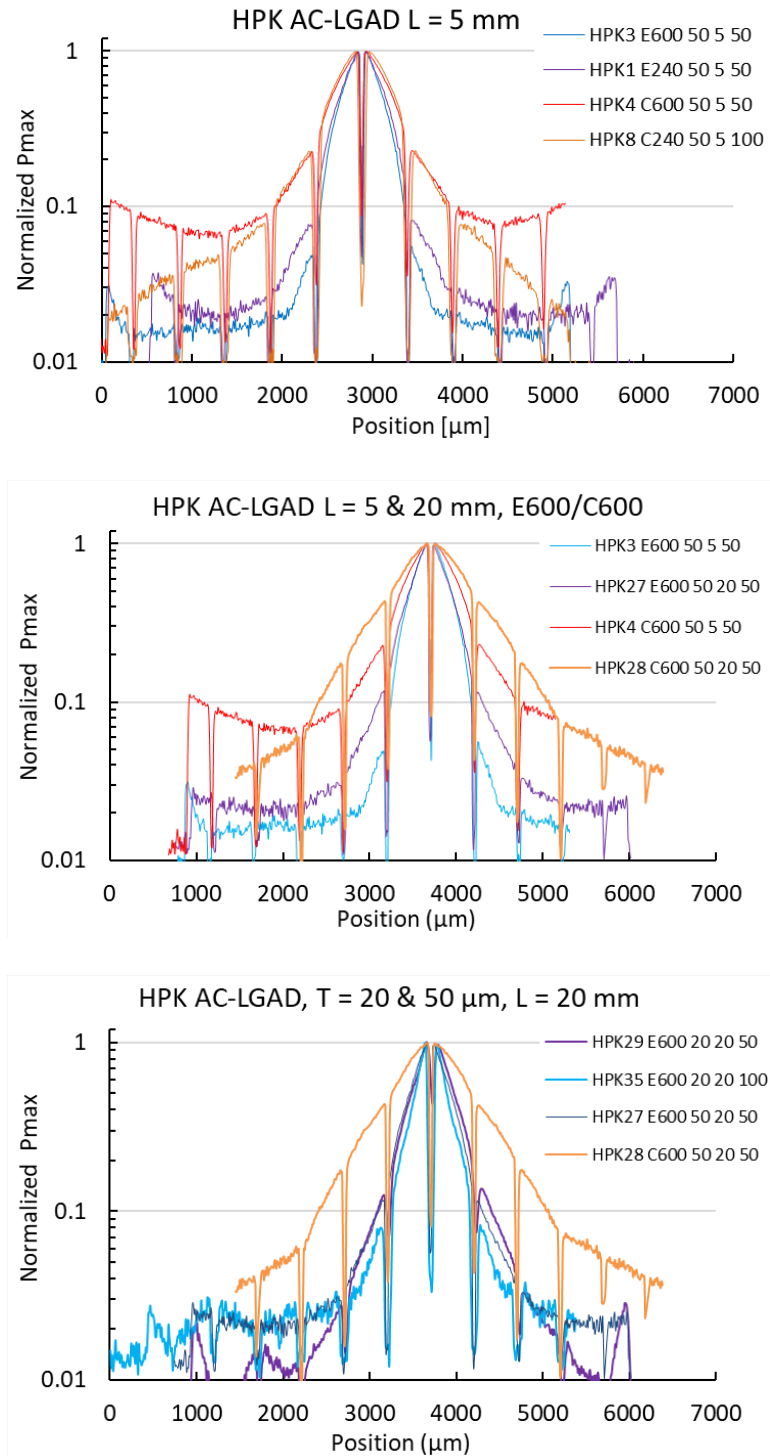

Fig. 4 Normalized Pmax distributions: TOP the four AC-LGADs of Fig. 2 RIGHT with strip length L = 5 mm; MIDDLE two E600 and C600 Type sensors each with length of 5 mm and 20 mm, respectively, BOTTOM E600 and C600 Type sensors with strip length L = 20 mm and bulk thickness T = 20 μm and 50 μm, respectively (Legends: ID#, Type, Bulk Thickness T, Strip Length L, Strip Width W).

The strip length influences the sharing to the next neighbor. Fig.4 MIDDLE shows the Pmax distributions for sensors with 5 mm and 20 mm length. The short- and long-distance sharing to the next neighbor strips is larger for the E600 and C600 sensors with strip length L = 20 mm than for ones with L = 5 mm. How the sharing changes with the thickness of the bulk T is shown in Fig 4 BOTTOM: a thinner E600 sensor with T = 20 µm has a better suppression of long-distance sharing than the one with T = 50 µm. In addition, the thinner detector with narrow metal width W = 50 µm has less long-range sharing than the one with W = 100 µm. As before, the C600 sensor with T = 50 shows large sharing diminishing at large distance.

### 3.3 Position Resolution

The TCT laser scans are being used to evaluate the positional jitter across two strips, which can be used to determine the position resolution when combined with possible constant position resolution errors. It depends on both the high precision position of the laser scan and on the normalized Pmax distributions between neighboring pairs of strips as shown in Fig. 5 for two AC-LGAD with 5 mm long strips, one a C240 (LEFT) and the other an E600 (RIGHT), respectively. The fraction Frac [13] is defined as

$$Frac = \frac{Pmax(1)}{[Pmax(1) + Pmax(2)]}$$

and is shown as function of the position in Fig. 6 LEFT. Both the inter-strip region for Pmax and the Frac distributions are different for E- and C-Types. The fraction shows a close to linear behavior between deep "notches" at the location of the strip centers, and this region shown in Fig. 6 RIGHT for all 10 sensors is used to calculate the slope ($\frac{dFrac}{dPos}$). The slope of fraction vs. position is very similar separately: for E-Type (slope = 0.0019 1/µ) and C-Type sensors (slope =0.0011 1/µ). The linear slope for the C-Type sensors being a factor 1.7 smaller than that for the E-Type sensors covers a much smaller position range, which means that the particle position needs to be calculated for a large part using the 2$^{nd}$ neighbor, which has a worse positional jitter due to smaller slope and smaller Pmax values (see below).

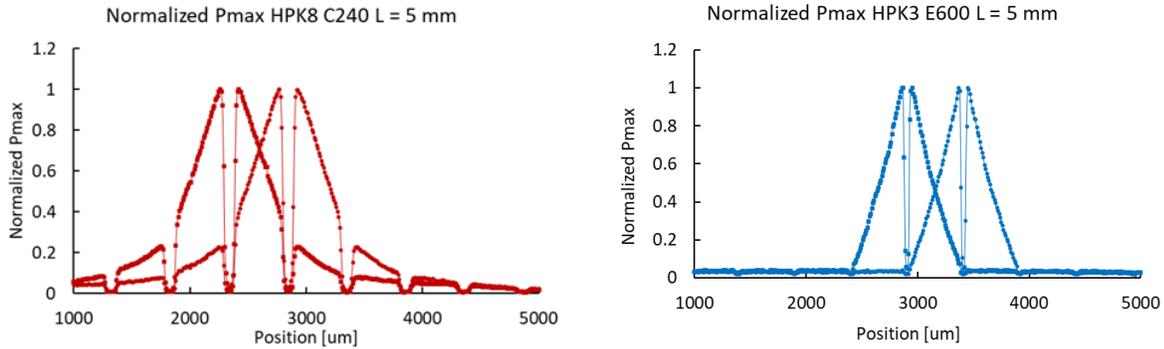

Fig. 5 Normalized Pmax distribution of two neighboring strips of LEFT C240 and RIGHT E600 sensors

The positional jitter $\sigma_J(Pos)$ can be derived from the fraction method in the linear region:

$$\sigma_J(Pos) = \sqrt{2}\left(\frac{dPos}{dFrac}\right)\frac{1}{S/N} \quad .$$

It is calculated from the inverse of the Frac slope ($dFrac/dPos$) , i.e. ($dPos/dFrac$), and uses the fact that the sum S = Pmax(1)+Pmax(2) is constant, and the errors in the two channels ( = electric noise N of Pmax) are identical. It is shown in Fig. 7 as a function of S/N. Like the fraction slope, the positional jitter falls into two groups depending on the resistance of the $N^+$ layer, independent of the coupling capacitance.

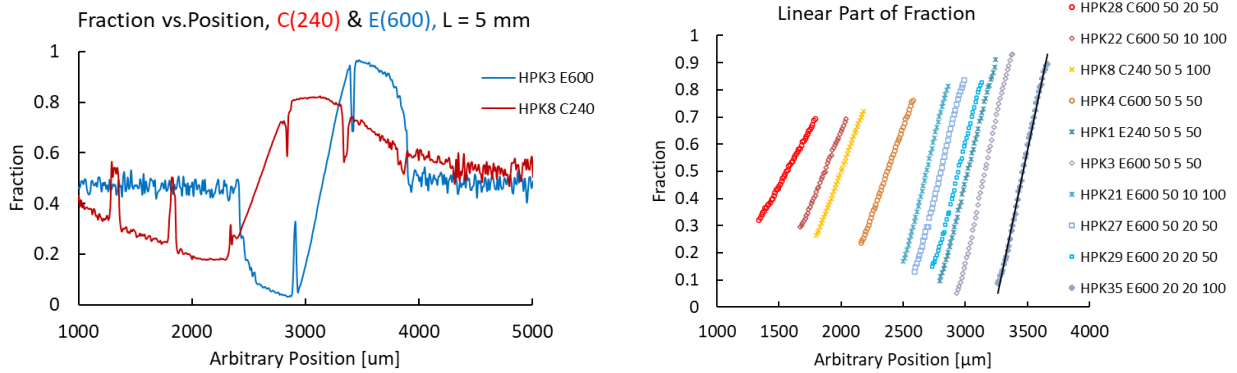

Fig. 6 Dependence of the fraction on the (arbitrary) position: LEFT for the sensors of Fig. 5, RIGHT for all sensors in the region of a linear slope, with two distinct groups in slope (C-Type red/yellow, E-Type blue) (Legends: ID#, Type, Thickness T, Strip Length L, Strip Width W)

The advantage of the E-Type sensors is shown by the better positional jitter compared to C-Type for the same signal height. For S/N ≥ 100 a positional jitter of $\sigma_J(Pos) \leq 8\ \mu m$ is reached for the E-Types independently of strip length, bulk thickness, and metal width. There is an outlier, HPK28, which is a 20 mm long C600 sensor shown in Fig.4 BOTTOM where the large sharing causes an inferior positional jitter, in addition to leading to a reduced Pmax as seen in Fig. 3.

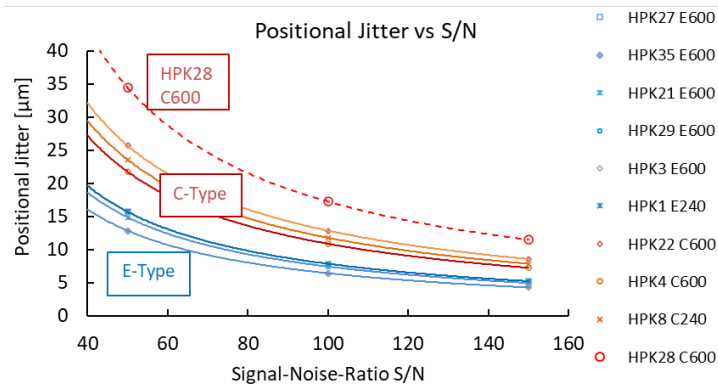

Fig. 7 Positional jitter vs. signal-to-noise ratio S/N for the 500 um pitch strip sensors.

4. **Conclusions**

In AC-LGADs, signal sharing needs to be high between neighboring strips for good position resolution and low for far-distant strips to reduce background signals. Using IR laser TCT, we have investigated the sharing in HPK sensors with variations of the parameters governing the sharing, i.e. the sheet resistance of the N$^+$ layer and the capacitance of the dielectric. Sensors with different layout of the metal strips were measured. The parameter choice "E600" gives good performance up to a strip length of 20 mm. It combines high sheet resistance of the N$^+$ layer with large coupling capacitance (thin di-electric layer). It maximizes the pulse height of the signal, leading to lower operating bias voltage, maximizes the amount of next neighbor sharing for good positional jitter and reduces the long-distance signal pick-up to the % level. A bulk thickness of 20 μm has the lowest long-distance pick-up. Assuming a realistic signal-to-noise ratio, the high resistance E-type sensor promises an excellent positional jitter of about 10 μm on a strip pitch of 500 μm for all tested strip parameters.

## 5. Acknowledgements

We acknowledge the collaboration with the KEK group (K. Nakamura et al.), the FNAL group (A. Apresyan at al.), the BNL group (A. Tricoli et al.), and the University of Illinois Chicago group (Z. Ye et al.). H.F.W.S. wants to thank the local organizing committee for the perfect organization and the very enjoyable atmosphere of HSTD13.
This work was supported by the United States Department of Energy grant DE-FG02-04ER41286, and by the U.S.-Japan Science and Technology Cooperation Program in High Energy Physics.

[15] Standa 8MT173
https://www.standa.lt/products/catalog/motorised_positioners?item=59&prod=motorized_translation_stages